# Asteroid (3200) Phaethon: results of polarimetric, photometric, and spectral observations

N. N. Kiselev,[1,*] V. K. Rosenbush,[2,3] D. Petrov,[1] I. V. Luk'yanyk,[3] O. V. Ivanova,[2,3,4] N. V. Pit,[1] K. A. Antoniuk,[1] V. L. Afanasiev[†]

[1] *Crimean Astrophysical Observatory, 298409 Nauchnij, Crimea*
[2] *Main Astronomical Observatory, National Academy of Sciences, UA-03143 Kyiv, Ukraine*
[3] *Taras Shevchenko National University of Kyiv, Astronomical Observatory,* UA-04053 Kyiv, Ukraine
[4] *Astronomical Institute, Slovak Academy of Sciences, SK-05960 Tatranská Lomnica, Slovak Republic*

[*] Corresponding Author. E-mail address: kiselevnn42@gmail.com

[†] V. L. Afanasiev passed away on December 21, 2020.

**ABSTRACT**

We present results of polarimetric, photometric, and spectral observations of the near-Earth asteroid (3200) Phaethon carried out at the 6-m BTA telescope of the Special Astrophysical Observatory and the 2.6-m and 1.25-m telescopes of the Crimean Astrophysical Observatory over a wide range of phase angles during its close approach to the Earth at the end of 2017 ($\alpha = 19° - 135°$) and in 2020 at $\alpha = 52.2°$. Using our and other available in literature data, we found that the maximum degree of linear polarization of Phaethon in the V band is $P_{max} = (45 \pm 1)\%$ at the phase angle $\alpha_{max} = 124.0° \pm 0.4°$, whereas the inversion angle $\alpha_{inv} = 21.4° \pm 1.2°$ and polarimetric slope is $h = (0.326 \pm 0.027)$ %/deg. Using the dependence "polarimetric slope – albedo," we have found the geometric albedo of asteroid Phaethon to be $p_v = 0.060 \pm 0.005$. This value falls into the lower range of albedo values for asteroids determined by different methods. The mean color indices $U–B = 0.207^m \pm 0.053^m$ and $B–V = 0.639^m \pm 0.054^m$ of the asteroid are derived at heliocentric and geocentric distances 1.077 au and 0.102 au, respectively, and phase angle $\alpha = 23.78°$. The absolute magnitude of Phaethon is $V(1,1,0) = 14.505^m \pm 0.059^m$. The effective diameter of Phaethon is estimated from obtained absolute magnitude and geometrical albedo, it is equal to $6.8 \pm 0.3$ km. The best fit to the observed polarimetric data was obtained with the Sh-matrix model of conjugated random Gaussian particles composed of Mg-rich silicate (90%) and amorphous carbon (10%).

# 1. Introduction

Potentially hazardous Apollo asteroid (3200) Phaethon (1983 TB) (hereafter Phaethon), discovered on October 11, 1983 by NASA's Infrared Astronomical Satellite (IRAS), is one of the most enigmatic objects among the known asteroids. It is classified as an asteroid but dynamically it is associated with the Geminid meteoroid stream that assumes that its source (Whipple 1983; Gustafson 1989; Williams & Wu 1993) experiences significant mass loss during its previous perihelion passages. Indeed, observations around perihelion in 2009 and 2012 obtained by the Solar and Terrestrial Relations Observatory (STEREO) revealed a strong brightening of Phaethon (Jewitt & Li 2010; Li & Jewitt 2013) and subsequent development of a comet-like tail (Jewitt, Li & Agarwal, 2013). This could suggest that Phaethon is an extinct or dormant comet, sometimes called a rock comet (Jewitt & Li 2010). However, as it was shown by de León et al. (2010) on the base of spectroscopic and dynamic data and by Alí-Lagoa et al. (2013, 2016) on the IR data, Phaethon could be a fragment of asteroid (2) Pallas with a "rock comet" orbit. Its unusual orbit has a high inclination ($i = 22.21°$), a very low perihelion distance ($q = 0.14$ au), an unusually high eccentricity ($e = 0.89$), and aphelion distance $a = 2.40$ au, i.e., it crosses the orbits of 4 planets; Mercury, Venus, Earth, and Mars. Despite the high eccentricity, the Tisserand parameter of Phaethon, equal to ~4.5, is asteroidal, not cometary. Due to the low perihelion, Phaethon is not only a near-Earth asteroid (NEA) but also a near-Sun asteroid (NSA).

Phaethon is one of the largest NEA. Up to the most recent apparition in 2017, its effective diameter was defined as $D = 5.1 \pm 0.2$ km (Hanuš et al. 2016). According to radar images of this asteroid obtained with the Arecibo planetary radar system during the 2017 apparition, Phaethon has a spheroidal shape more than 6 km (the best-fit spherical diameter for Phaethon is about 6.2 km) in diameter at the equator with several distinguishable surface features, among them a dark feature near one of the poles and about 1 km-sized depression-like feature near the equator (Taylor et al. 2019). From photometric observations, a rotation period was found to be 3.604 h (Kim et al. 2018; Hanuš et al. 2018).

Phaethon was classified as a F-type asteroid by Tholen (1985a,b) and a B-type asteroid by Green, Meadow & Davies (1985) and Bus & Binzel (2002). Subsequently, Licandro et al. (2007) revealed an absorption band in the UV spectra of Phaethon that is inconsistent with the F class. However, the recent spectroscopic observations of the asteroid (see, e.g., Kareta et al. 2018; Lazzarin et al. 2019) did not detect a near-UV absorption in the UV spectra of Phaethon. The spectra of Phaethon exhibit the mostly blue spectral slope in the VIS-NIR region, which is bluer than any other small bodies in the solar system, consistent with B-type asteroids. Spectral gradient is time-variable, thus suggesting there exists inhomogeneity in surface properties of Phaethon (Lazzarin et al. 2019; Ohtsuka et al. 2020). In contrary, Lee et al. (2019) concluded that southern hemisphere of Phaethon has a homogeneous surface and there was no surface inhomogeneity in latitude direction. Thus, the issue of the surface homogeneity (or inhomogeneity) of Phaethon remains open.

The visible geometric albedo of asteroid Phaethon obtained by different authors is varied within the large range (see Table 3): from $p_v = 0.16 – 0.17$ (Veeder, Kowal & Matson 1984; Usui et al. 2011) to $p_v = 0.05 – 0.06$ (Devogèle et al. 2018; Taylor et al. 2019). The JPL Small Body Database (JPL777) has taken the geometric albedo of Phaethon to be $0.1066 \pm 0.011$.

According to Jewitt & Li (2010), Phaethon is essentially a rock comet, a rare type of object that releases rock particles due to a thermal fracture at small perihelion distances. Extreme solar heating (up to 1000 K) at the time of very small perihelion distance and rapid temperature variations on the surface can cause thermal fracturing and/or desiccation cracking of the surface material of the asteroid, thus, producing dust particles (Delbo et al. 2014). Carbonaceous species, phyllosilicates (clays) or other hydrated minerals have been suggested as components of dust particles at the Phaethon surface (Licandro et al. 2007). According to Delbo et al. (2014), thermal fragmentation induced by the diurnal temperature variations breaks up rocks larger than a few centimeters and releases dust that is carried away by the radiation pressure of sunlight. Outgoing dust particles are about a millimeter across, approximately the size of the Geminids. However, observations with Hubble Space Telescope during the 2017 apparition revealed neither coma nor dust ejection from Phaethon at heliocentric distance around 1 au (Ye et al. 2019; Devogèle et al. 2020).

More than twenty years ago, in our paper on polarimetry of asteroid (2100) Ra-Shalom (Kiselev, Rosenbush & Jockers 1999), we wrote that the behavior of polarization of low-albedo asteroids at large phase angles was not known. However, recently the situation has changed. In 2015, Kuroda et al. (2018) measured the linear polarization degree of the low-albedo Cb-type asteroid NEA (152679) 1998 KU$_2$ at a large phase angle, which turned out the highest of any known atmosphereless solar system bodies at similar phase angles: $P = 44.6 \pm 0.5\%$ in the R band at a phase angle $\alpha = 81.0°$. As for asteroid Phaethon, its polarization was first measured at phase angle $\alpha = 22.9°$, near inversion angle, in December 14, 2012 by Fornasier et al. (2006). And only in September-November 2016, Ito et al. (2018) conducted detailed polarimetric observations of this asteroid over a wide range of phase angles, from 106.5° to 33.0°. These observations allowed them to obtain the phase-angle dependence of polarization for Phaethon and reveal that it has extremely high linear polarization: $P = 50.0 \pm 1.1\%$ at $\alpha = 106.5°$, and its maximum is probably located at much larger $\alpha$ than 106.5°, and then $P_{max}$ of Phaethon is substantially larger than 50%. The strong polarization implies that Phaethon's geometric albedo is lower than the current estimate obtained through radiometric observations (see Table 3). During the 2017 apparition, a multi-color polarization curve of asteroid Phaethon has been obtained at the phase angles ranging from 36° to 116° by Devogèle et al. (2018) and Borisov et al. (2018). The authors concluded that the covered interval of phase angles was not sufficiently extended to derive a firm determination of the $P_{max}$ parameter, which supposedly occurred at phase angle around 130° and reached more than 45%. Shinnaka et al. (2018) also obtained the phase-polarization curve for Phaethon for a wide range of $\alpha$ from 19° to 114° with $P_{max}$ not less than 42.4% at $\alpha_{max}$ not less than 114.3°. In the 2017 apparition, polarization of Phaethon was also measured by Zheltobryukhov et al. (2018) at phase angles 57.9° and 73.2° and Okazaki et al. (2020) at phase angles 33.4° and 57.9°. All listed measurements of polarization were performed in the standard Johnson-Cousins photometric system. The derived values of linear polarization appeared to be significantly larger than those for the moderate albedo asteroids.

The unprecedented close approach of Phaethon to the Earth on December 16, 2017 at the geocentric distance of 0.069 au provided us with a good opportunity for comprehensive monitoring of the asteroid up to large phase angles. The observations of the asteroid were conducted at the 6-m telescope of the Special Astronomical Observatory (SAO) and 2.6-m and 1.25-m of the Crimean Astrophysical Observatory (CrAO). The main goal of these observations was to obtain a complete phase-angle dependence of polarization, including the polarization maximum and the phase angle at which it occurs. This would allow one to directly determine albedo and to model characteristics of the regolith particles. The second aim of our observations was to define the variation of the polarization with wavelength. Besides, Phaethon and (101955) Bennu, which is also a B-type asteroid, are the targets of two space missions, the JAXA Destiny⁺ and NASA OSIRIS-Rex, respectively. Therefore, another goal of our work was to obtain as much information as possible about the physical nature of these asteroids and extend previous researches by studying their polarimetric properties.

We present characteristics of asteroid Phaethon derived from polarimetric, photometric, and spectral observations. In Section 2, we briefly describe observations of Phaethon and the data reduction. The results of polarimetric observations, including the parameters of the phase-angle dependence of polarization and albedo of Phaethon are described in Section 3. Modeling of the phase-angle dependence of polarization for Phaethon is put forward in Section 4. In Section 5, we show the results of the UBV photometry of Phaethon. The visible reflectance spectra of Phaethon are discussed in Section 6. Finally, in Section 7 we summarize our findings.

2. Observations, instruments, and data reductions

We observed asteroid Phaethon at three telescopes from November 27 to December 26, 2017, while it was approaching perihelion and was near its closest approach to the Earth (December 16, 2017). At that time, the heliocentric distance ($r$) gradually decreased from 1.305 to 0.840 au, while the geocentric distance ($\Delta$) and the phase angle ($\alpha$) of Phaethon varied widely: $\Delta$ decreased from 0.382 to 0.0768 au and then increased to 0.191 au, $\alpha$ changed from 28.7° to 19.2°, and then increased to 134.9°. We also observed asteroid Phaethon on October 16, 2020. The CCD spectropolarimetry and imaging polarimetry were performed at the 6-m telescope of the SAO Aperture polarimetry and imaging photometry were carried out at the 2.6-m and 1.25-m of the CrAO.

*2.1. Spectropolarimetry and imaging polarimetry*

The multi-mode focal reducers SCORPIO-1 (Spectral Camera with Optical Reducer for Photometrical and Interferometrical Observations) (Afanasiev & Moiseev 2005) and SCORPIO-2 (Afanasiev & Moiseev 2011; Afanasiev & Amirkhanyan 2012) installed in the primary focus ($f/4$) of the 6-m BTA telescope of the SAO were used for observations of asteroid Phaethon in the long-slit spectropolarimetry (hereafter called SpPol) and imaging linear polarimetry (ImaLP) modes during the period of observations from November 27 to December 26, 2017 and on October 16, 2020 (Table 1). The telescope tracked the motion of the fast moving asteroid to compensate its proper velocity during the exposures.

The low-resolution linear spectropolarimetry of Phaethon was conducted with the SCORPIO-2 instrument from November 27 to December 13, 2017. In the spectropolarimetric mode of the instrument, an achromatic half-wave phase plate and the Wollaston prism were installed before a grism. The Wollaston analyzer splits images of the ordinary $I_o(\lambda)$ and extraordinary $I_e(\lambda)$ rays by 5°. For fixed positions of the $\lambda/2$ phase plate 0°, 45°, 22.5°, and 67.5°, a series of pairs of spectra $I_o(\lambda)$ and $I_e(\lambda)$ in mutually perpendicular polarization planes was obtained at the exit of the spectrograph. The spectral grating VPHG940@600 provided an effective wavelength range within $\lambda$4100−8100 Å with the 5 Å spectral resolution for the slit width 0.47". CCD E2V 42-90 matrix used as the detector with binning $2 \times 4$ which provided image scale of 0.357" × 0.714". The exposure for each measurement was chosen to allow getting the required signal-to-noise (S/N) ratio (usually about 200 after co-adding images) and to minimize the effect of the Earth's atmosphere (depolarization). Image of the spectra of a continuous-spectrum lamp was used for correction of flat field, which was obtained for each angle of the phase plate. The spectrograph slit, with the height of 6' and width of 0.47", was oriented along the asteroid's velocity vector. During the observations, the amplitude of the atmospheric differential refraction did not exceed the width of the used slit. In further analysis, the true values of the Stokes parameters $I(\lambda)$, $Q(\lambda)$, and $U(\lambda)$ were obtained, from which the degree of linear polarization $P(\lambda)$ and the position angle of the plane of polarization $\theta(\lambda)$ were calculated. Note that a large advantage of the design of the instrument SCORPIO-2 is that it does not introduce any significant instrumental polarization along the slit height. To calibrate the polarization tract of the spectrograph, we observed the polarimetric standards from the lists of Hsu & Breger (1982) and Schmidt, Elston & Lupie (1992) every night.

The primary reductions of the spectropolarimetric observations were performed with the specialized software packages in the IDL environment developed in the SAO RAS (Afanasiev & Amirkhanyan 2012), which included the bias subtraction, the flat field correction, geometrical correction along the slit, the sky background subtraction, spectral sensitivity of the instrument, the spectral wavelength calibration, the presentation of data with uniform scale spacing along the wavelengths, extraction of the spectra from the images, and the calculation of the Stokes parameters. To estimate the level of sky background, we obtained the night sky spectrum at the same positions of the analyzer, which was subtracted from the spectra of the asteroid. The polarization parameters $Q(\lambda)$ and $U(\lambda)$ were robustly estimated as average for the entire cycle of measurements. To correctly determine the errors, we computed robust estimations of the Stokes parameters $Q(\lambda)$ and $U(\lambda)$ and standard deviations at $3\sigma$ (the 3-sigma criterion) in the spectral ranges of 70 Å: values more than 3-sigma were ignored. This type of measurements gives a good estimate of statistical errors of the measured parameters. The final polarization values in the broadband B($\lambda$4420/890 Å), V($\lambda$5360/820 Å), and R($\lambda$6450/1550 Å) filters were calculated by integrating $P(\lambda)$ multiplied by the standard curve of the filter passband (hereinafter we indicate the central wavelength $\lambda_0$ and *FWHM*).

On December 18–26, 2017, the SCORPIO-1 instrument was used in a mode of imaging polarimetry for the measurement of linear polarization of Phaethon. The wide-band V filter ($\lambda$5360/820 Å) was used. For obtaining the direct images, the large-format CCD E2V 42-90 matrix of 2048 × 2048 pixels with a pixel size of 13.5 × 13.5 μm was used as a detector. This matrix yields a full field of view is 6.1' × 6.1' with a pixel scale of 0.18 arcsec/px. A Savart plate that separates the beams in two mutually perpendicular planes of polarization was used as the polarization analyzer which can be turned around the optical axis through 45°. A stepping motor was used to put the analyzer in and withdraw from the beam and to turn it. Several cycles of measurements were carried out during the observations, when we successively recorded the image of the asteroid for pairs of angles. To improve the S/N ratio of the measured signal, on-chip binning 2 × 2 was applied to all observed images.

The reduction process of the raw images obtained with the SCORPIO-1 included subtracting a bias image, dividing by a bias-subtracted flat-field, and removing traces of cosmic rays. The bias was removed by subtracting an averaged frame with zero exposure time. The twilight sky was exposed to provide flat-field corrections for the non-uniform sensitivity of the CCD matrix. Removal of the traces of cosmic rays was done at the final stage of the reduction via a robust parameter estimates for reducing a bias caused by outliers (Fujisawa 2013). The sky background level was determined on the basis of estimations from a histogram of counts in the image measured in regions outside of the asteroid and free of faint stars. To increase the S/N ratio, we added together all images of the asteroid taken during the night and summed using a robust averaging method (Rousseeuw & Bassett 1990). To avoid artifacts, the images of the asteroid were centered with a precision of 0.1 px using the maximum brightness of the asteroid.

A detailed description of the SCORPIO-1 and SCORPIO-2 and the technique of observations, the image processing, reduction, errors estimation, and the method of calculation of polarization parameters in different modes of the instruments are also described in (Kiselev et al. 2013; Afanasiev, Rosenbush & Kiselev 2014; Ivanova et al. 2017; Rosenbush et al. 2020).

*2.2. Aperture polarimetry and photometry*

On December 8–13, 2017, a five-color UBVRI photopolarimeter (Piirola, 1988) of the 1.25-m AZT-11 telescope and a single-channel photometer-polarimeter of the 2.6-m Shajn telescope of the CrAO (Shakhovskoy & Efimov 1972; Kolesnikov et al. 2016) were used for the measurements of linear polarization of asteroid Phaethon at phase angles 19°–62°.

In the UBVRI aperture photopolarimeter mounted at the 1.25-m AZT-11 CrAO telescope, the polarizer is a plane parallel calcite slabe that produces in a focal plane of the telescope two orthogonally polarized intensities (in two diaphragms) which are measured by single photomultiplier for each spectral band using chopping techniques. The measured radiation flux is modulated with a half-wave phase plate discretely rotating with a step of 22.5°. The complete cycle of one measurement contains eight exposures at eight different positions of the phase plate. In each of the diaphragms, two beams from the sky background with orthogonal directions of oscillations of electric vector overlap. Because of this, there is no need to account for the polarization of the sky background, and only the background brightness is taken into account. This is one of the advantages of the observational technique that uses the five-color polarimeter. The photometric system is realized with four dichroic filters which separate the light into five spectral regions, i.e., the instrument provides simultaneous measurements of polarization in the UBVRI bands ($\lambda$3600, 4400, 5300, 6900, and 8300 Å, respectively). The passbands are close to the standard UBV (Johnson) and RI (Cousins) systems. Due to some technical problems, the polarimetric observations of Phaethon at the 1.25-m telescope were carried out only in the VRI filters. The observations fulfilled with this polarimeter were processed according to the program developed by Berdyugin & Shakhovskoy (1993).

A single-channel photometer-polarimeter of the 2.6-m Shajn telescope, operating with the photoelectronic multiplier as a radiation detector in the photon-counting mode, is designed to quasi-simultaneously measure all four Stokes parameters *I*, *Q*, *U*, and *V*. The polarimetric unit of the instrument includes the achromatic five-component quarter-wave phase plate as a polarizer, continuously rotating with the rate of 33 Hz, and the Glan prism as an analyzer placed behind it. To control the rotation angle of the modulator, the unit contains optical-mechanical assembly. The recording system accumulates the impulses over the time intervals corresponding to the 22.5° angles of the retarder rotation, i.e., by eight pulse counters. We used the 15 arcsec circular diaphragm for all observations. The sky background was measured at the beginning and at the end of the observations of Phaethon and between the series of measurements of the asteroid. The values of the intensity of the background were interpolated and taken into consideration for each of the measurements of the object. The B($\lambda$4340/800 Å) and V($\lambda$5450/800 Å) bands of Johnson photometric systems and the WR($\lambda$6360/2630 Å) wide-red filter were used for observations.

The instrumental polarization was determined from many observations of unpolarized standard stars selected from Serkowsky (1974). The zero point of the position angle was determined from the measurement of highly polarized standard stars (Serkowsky 1974; Hsu & Breger 1982). We found that the instrumental polarization of the 2.6-m telescope is less than 0.2% in the B, V, and WR filters, while the instrumental polarization of the 1.25-m telescope was $(1.56 \pm 0.02)$% in the V filter, $(0.63 \pm 0.02)$% in the R filter, and $(0.39 \pm 0.03)$% in the I filter. This instrumental polarization was removed from the measured polarization of Phaeton using the normalized Stokes parameters $q$ and $u$. The zero-point of the position angle of the polarization plane was stable within ~1° for both telescopes during our observational runs.

The degree of linear polarization was calculated from the dimensionless Stokes parameters $u = U/I$ and $q = Q/I$ acquired in the observations of Phaethon for each of the dates as $P = \sqrt{u^2 + q^2}$. The position angle of the polarization plane in the instrumental coordinate system was calculated according to the formula $\theta = 1/2 \arctan(u/q)$. The position angle $\theta_r$ of the polarization plane relative to the plane orthogonal to the scattering plane was found from $\theta_r = \theta - (\varphi \pm 90°)$, where $\varphi$ is the position angle of the scattering plane, and the sign in bracket is chosen to provide $0 \leq \varphi \pm 90° < 180°$ (Chernova, Kiselev & Jockers 1993). The error in determining the polarization degree was estimated by two ways: from the statistics of the accumulated impulses and from the dispersion of the Stokes parameters calculated for each of the series during the night. The larger of these two errors was considered as the measurement error of the polarization degree (Shakhovskoi & Efimov 1972). In both cases, the error in determining the angle of the polarization plane was found with the formula $\sigma_\theta = 28.65 \sigma_P / P$.

We used good photometric conditions on December 12, 2017 to estimate the $V$ magnitudes and color indices $U–B$ and $B–V$ of Phaethon, using polarimetric observations at the 1.25-m telescope. Photometric calibration was performed using the polarimetric standard star HD 25914 ($V = 8.07^m$, $U–B = –0.25^m$, $B–V = 0.49^m$), which, like Phaethon, was observed in the polarimetric mode.

The viewing geometry and log of observations of asteroid Phaethon are presented in Table 1. There we list the date of observation (the middle of integration times is taken, UT), the heliocentric ($r$) and geocentric ($\Delta$) distances, the phase angle (Sun-Comet-Earth angle) ($\alpha$), the position angle of the scattering plane ($\varphi$), the filter or grism, the mode of the observation where SpPol is spectropolarimetry, ApPol is aperture polarimetry, Phot is photometry, ImaPol is the imaging polarimetry, number of cycles of exposures obtained in one night ($N$) and the exposure time during one cycle ($T_{exp}$), and the telescope/observatory. The apparent motion of Phaethon was very fast, so we chose short exposure times to keep the object in the 20 and 15 arcsec diaphragms of the polarimeters of the 1.25-m and 2.6-m telescopes respectively. Individual exposure time was chosen depending on the observation mode, the apparent magnitude of the asteroid, filter, and observation conditions.

On December 26.625, 2017, when the phase angle was 134.9°, observations of Phaethon were made at zenith distance $Z = 76°$ at the airmass of 4.2. The use of the Savart plate makes it possible to obtain simultaneously images of the asteroid in two mutually perpendicular polarization planes. Consequently, each of the polarization parameters $Q$ and $U$ is practically independent on the observation conditions (e.g., on the airmass). The reduction of several cycles of asteroid measurements for pairs of the Savart plate angles makes it possible to reliably determine the $Q$ and $U$ parameters for the average observation moment. Thus, according to our analysis, the large air mass had only a minor effect on our retrieved data.

**Table 1**
The viewing geometry and observation log of asteroid (3200) Phaethon.

| Date (UT) | $r$ (au) | $\Delta$ (au) | $\alpha$ (deg) | $\varphi$ (deg) | Filter/grism | Mode | $N \times T_{exp}$ (sec) | Telescope/ Observatory |
|---|---|---|---|---|---|---|---|---|
| 2017 | | | | | | | | |
| Nov 27.121 | 1.305 | 0.382 | 28.67 | 256.9 | VPHG940@600 | SpPol | 7 × 60 | 6.0/SAO |
| Nov 28.005 | 1.293 | 0.365 | 28.22 | 255.7 | VPHG940@600 | SpPol | 5 × 120 | 6.0/SAO |
| Dec 08.758 | 1.141 | 0.167 | 19.79 | 212.1 | V, R, I | ApPol | 27 × 120 | 1.25/CrAO |
| Dec 09.001 | 1.137 | 0.163 | 19.61 | 209.4 | VPHG940@600 | SpPol | 10 × 120 | 6.0/SAO |
| Dec 09.754 | 1.126 | 0.150 | 19.21 | 200.2 | VPHG940@600 | SpPol | 5 × 120 | 6.0/SAO |
| Dec 12.775 | 1.079 | 0.104 | 23.25 | 145.6 | VPHG940@600 | SpPol | 11 × 120 | 6.0/SAO |
| Dec 12.901 | 1.077 | 0.102 | 23.78 | 142.9 | U, B, V, R, I | ApPol/Phot | 45 × 120 | 1.25/CrAO |
| Dec 13.734 | 1.064 | 0.091 | 28.34 | 125.5 | VPHG940@600 | SpPol | 10 × 60 | 6.0/SAO |
| Dec 13.734 | 1.064 | 0.091 | 28.34 | 125.5 | V, R, I | ApPol | 44 × 120 | 1.25/CrAO |
| Dec 15.688[a] | 1.032 | 0.073 | 47.39 | 89.4 | WR | ApPol | 53 × 16 | 2.6/CrAO |
| Dec 15.728[a] | 1.031 | 0.072 | 47.91 | 88.8 | V | ApPol | 50 × 16 | 2.6/CrAO |
| Dec 15.763[a] | 1.031 | 0.072 | 48.35 | 88.3 | B | ApPol | 50 × 16 | 2.6/CrAO |
| Dec 16.720[a] | 1.015 | 0.069 | 61.65 | 76.2 | WR | ApPol | 22 × 12 | 2.6/CrAO |
| Dec 18.799 | 0.98 | 0.077 | 90.63 | 65.1 | V | ImaLP | 8 × 2 × 60 | 6.0/SAO |
| Dec 19.658 | 0.96 | 0.085 | 100.1 | 64.7 | V | ImaLP | 5 × 2 × 60 | 6.0/SAO |
| Dec 26.625[b] | 0.84 | 0.191 | 134.9 | 77.6 | V | ImaLP | 2 × 260 | 6.0/SAO |
| 2020 | | | | | | | | |
| Oct 16.088 | 1.212 | 1.018 | 52.2 | 286.8 | V | ImaLP | 2 × 260 | 6.0/SAO |

[a] partially cloudy; [b] observations were made at zenith distance $Z = 76°$.

## 3. Results of polarimetric observations

In Table 2, we present the results of our polarimetric observations of asteroid Phaethon in the B, V, R, WR, and I filters. The table contains the observation time (UT), the filter, the phase angle $\alpha$, the position angle $\varphi$ of the scattering plane, the measured degree of linear polarization $P$ of Phaethon and its error $\sigma_P$, where $P$ is considered positive if the polarization plane is parallel to the scattering plane and negative if the polarization plane is perpendicular to the scattering plane, the position angle $\theta$ of the polarization plane in the equatorial coordinate system with its error $\sigma_\theta$, and the position angle $\theta_r$ of the polarization plane and polarization degree relative to the plane perpendicular to the scattering plane. Observation conditions in the SAO were poor, so the errors in the degree of polarization are large.

*3.1. Phase-angle dependence of polarization for Phaethon*

In Fig. 1, we present all available polarimetric data for asteroid Phaethon obtained in the period 2004–2020 with the B, V, R, and I filters. In addition to our measurements (Table 1), there are the data by Fornasier et al. (2006) at the phase angle $\alpha = 22.9°$, Devogèle et al. (2018) obtained at the Calern Observatory at $\alpha = 36.8° – 101.9°$ and the Rozhen Observatory at $\alpha = 47.5° – 116.2°$, Shinnaka et al. (2018) at $\alpha = 19.2° – 114.2°$, Okazaki et al. (2020) at $\alpha = 33.4° – 57.9°$, and Zheltobryukhov et al. (2018) at the phase angles 57.9° and 73.2°. Ito et al. (2018) observed Phaethon in the 2016 apparition at $\alpha = 106.5° – 33.0°$.

**Table 2**
Results of polarimetric observations of asteroid (3200) Phaethon.

| Date (UT) | α (deg) | φ (deg) | Filter | P ± σ_P (%) | θ ± σ_θ (deg) | θ_r (deg) | P_r (%) | Telescope (m) |
|---|---|---|---|---|---|---|---|---|
| 2017 | | | | | | | | |
| Nov 27.121 | 28.67 | 256.9 | V | 2.53 ± 0.39 | 153.0 ± 18.0 | –13.9 | 2.24 | 6.0 |
| 28.005 | 28.22 | 255.7 | V | 2.12 ± 0.28 | 161.0 ± 2.0 | –4.7 | 2.09 | 6.0 |
| Dec 08.758 | 19.79 | 212.1 | V | 0.47 ± 0.39 | 170.6 ± 19.9 | 48.5 | –0.06 | 1.25 |
| 08.758 | 19.79 | 212.1 | R | 0.25 ± 0.17 | 170.3 ± 17.3 | 48.2 | –0.03 | 1.25 |
| 08.758 | 19.79 | 212.1 | I | 0.10 ± 0.28 | 12.9 ± 34.7 | 70.8 | –0.08 | 1.25 |
| 09.001 | 19.61 | 209.4 | B | 0.83 ± 0.45 | 153.1 ± 15.6 | 33.7 | 0.32 | 6.0 |
| 09.001 | 19.61 | 209.4 | V | 0.33 ± 0.32 | 171.0 ± 12.0 | 51.6 | –0.08 | 6.0 |
| 09.001 | 19.61 | 209.4 | R | 0.83 ± 0.36 | 144.9 ± 12.6 | 25.5 | 0.52 | 6.0 |
| 09.754 | 19.21 | 200.2 | B | 0.86 ± 0.22 | 159.2 ± 7.5 | 49.0 | –0.12 | 6.0 |
| 09.754 | 19.21 | 200.2 | V | 0.73 ± 0.31 | 22.0 ± 5.0 | 91.8 | –0.73 | 6.0 |
| 09.754 | 19.21 | 200.2 | R | 1.06 ± 0.14 | 164.4 ± 3.8 | 54.3 | –0.34 | 6.0 |
| 12.775 | 23.25 | 145.6 | B | 1.92 ± 0.28 | 41.1 ± 4.2 | –14.5 | 1.68 | 6.0 |
| 12.775 | 23.25 | 145.6 | V | 1.02 ± 0.31 | 81.0 ± 6.0 | 25.4 | 0.64 | 6.0 |
| 12.775 | 23.25 | 145.6 | R | 1.66 ± 0.19 | 30.3 ± 3.3 | –25.3 | 1.06 | 6.0 |
| 12.901 | 23.78 | 142.9 | B | 2.85 ± 0.26 | 65.9 ± 2.3 | 13.0 | 2.56 | 1.25 |
| 12.901 | 23.78 | 142.9 | V | 0.80 ± 0.17 | 64.3 ± 6.1 | 11.4 | 0.74 | 1.25 |
| 12.901 | 23.78 | 142.9 | R | 1.14 ± 0.10 | 57.8 ± 2.6 | 4.9 | 1.12 | 1.25 |
| 12.901 | 23.78 | 142.9 | I | 1.37 ± 0.15 | 49.7 ± 3.1 | –3.2 | 1.36 | 1.25 |
| 13.734[a] | 28.34 | 125.5 | B | 2.41 ± 0.39 | 23.3 ± 4.7 | –12.2 | 2.19 | 6.0 |
| 13.734[a] | 28.34 | 125.5 | R | 2.05 ± 0.48 | 24.2 ± 6.7 | –11.3 | 1.89 | 6.0 |
| 13.734 | 28.34 | 125.52 | B | 3.15 ± 0.26 | 51.7 ± 1.6 | 16.2 | 2.66 | 1.25 |
| 13.734 | 28.34 | 125.5 | V | 2.34 ± 0.22 | 29.1 ± 2.6 | –6.4 | 2.28 | 1.25 |
| 13.734 | 28.34 | 125.5 | R | 2.46 ± 0.13 | 38.7 ± 1.5 | 3.2 | 2.44 | 1.25 |
| 13.734 | 28.34 | 125.5 | I | 2.83 ± 0.18 | 36.5 ± 1.9 | 1.0 | 2.83 | 1.25 |
| 15.688 | 47.39 | 89.4 | WR | 10.85 ± 0.74 | 4.3 ± 2.0 | 4.8 | 10.69 | 2.6 |
| 15.728 | 47.91 | 88.8 | V | 10.50 ± 0.85 | 2.9 ± 2.3 | 4.1 | 10.39 | 2.6 |
| 15.763 | 48.35 | 88.3 | B | 10.54 ± 1.40 | 5.3 ± 3.8 | 7.0 | 10.23 | 2.6 |
| 16.720 | 61.65 | 76.2 | WR | 18.01 ± 1.02 | 170.2 ± 1.6 | 4.0 | 17.84 | 2.6 |
| 18.799 | 90.63 | 65.1 | V | 33.2 ± 0.7 | 151 ± 2 | –4.1 | 32.9 | 6.0 |
| 19.658 | 100.1 | 64.7 | V | 37.3 ± 0.4 | 158 ± 3 | 3.3 | 37.0 | 6.0 |
| 26.625[b] | 134.9 | 77.6 | V | 32 ± 3 | (168) | – | 32 | 6.0 |
| 2020 | | | | | | | | |
| Oct. 16.088 | 52.21 | 286.8 | V | 12.68 ± 0.50 | 18.4 ± 1.1 | 1.6 | 12.66 | 6.0 |

[a] partially cloudy; [b] observations were made at zenith distance $Z = 76°$.

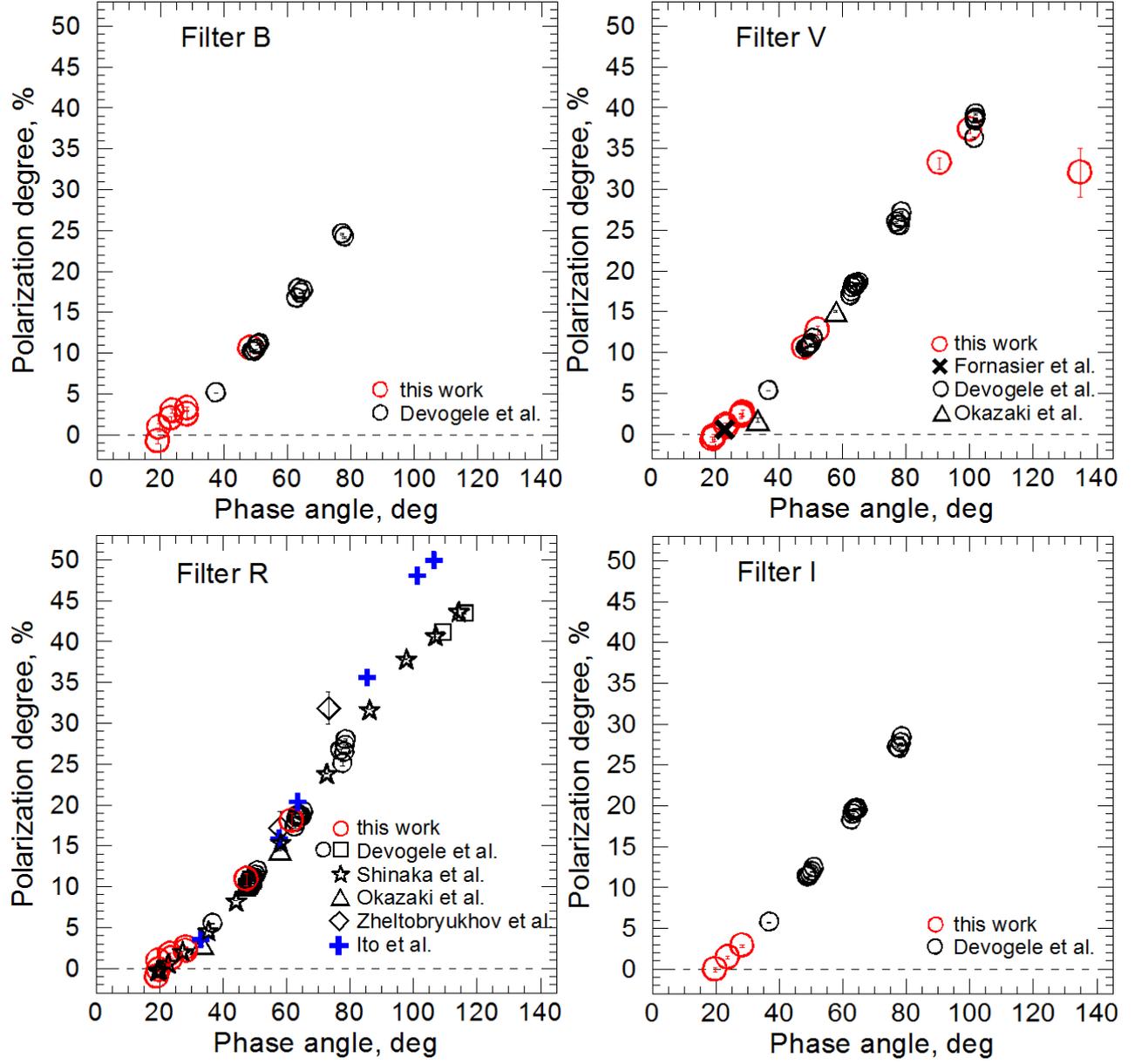

**Figure 1.** The polarization degree vs phase angle for asteroid (3200) Phaethon in the BVRI filters. Data are taken from Fornasier et al. (2006) (black oblique cross), Ito et al. (2018) (blue crosses), Devogele et al. (2018) (Calern Observatory – black circles; Rhozen Observatory – black squares), Shinnaka et al. (2018) (black asterisks), Okazaki et al. (2020) (black triangles), Zheltobryukhov et al. (2018) (black rhombuses), and present work (red circles).

To analyze the phase-angle dependence of polarization, the authors use either the observed degree of polarization $P$ and the position angle $\theta$ of the polarization plane in an equatorial coordinate system, or parameters $P_r$ and $\theta_r$ referred to the scattering plane, i.e. the plane that contains the incident and the scattered radiation. These parameters are interconnected according to the following expressions: $P_r(\alpha) = P(\alpha)\cos 2\theta_r$ and $\theta_r = \theta - (\varphi \pm 90)$, where $\varphi$ is the position angle of the scattering plane. All observations with $P > \sigma_P$, where $\sigma_P$ is the error in determining the polarization degree, show that $\theta_r$ is equal to about 90° or 0°, i.e. parallel or perpendicular to the scattering plane. However, in the case when the measured polarization $P \leq \sigma_P$, the uncertainty in the observed position angle of the polarization plane, defined by the expression $\sigma_\theta = 28.65\,\sigma_P/P$ or even $\sigma_\theta = \pi/12^{0.5}$ rad = 52° (Naghizadeh-Khouei & Clarke 1993), becomes large, and the values $\theta_r$ deviate significantly from 90° or 0°. Since always $|P_r| \leq P$, then $P_r$ becomes unrealistic. Therefore,

Kiselev & Petrov (2018) suggested to apply the following relations: $P_r = -P$ if $45° \leq \theta_r \leq 135°$ and $P_r = P$ if $-45° < \theta_r < 45°$. Thus, to construct polarization curve, we used the $P_r$ values taken from Devogèle et al. (2018) and Shinnaka et al. (2018), for which $P >> \sigma_P$, and the $P$ values from other measurements in which the error $\sigma_P$ in the polarization degree is quite large.

As can be seen in Fig. 1, all data obtained in the 2017 apparition show similar behaviour of polarization of Phaethon with the phase angle. However, it is more important to point out that the polarization of the asteroid measured in the 2016 apparition by Ito et al. (2018) at the large phase angles significantly differed from the data derived in 2017 at corresponding angles (see the panel for the R filter). According to the 2016 observations, Phaethon exhibited unusually high polarization, up to 50% at $\alpha = 106.5°$ in the R filter. The cause for this discrepancy could be the changing aspect angle, i.e., the angle between the line of sight and the axis of rotation of the asteroid, and, hence, different regions of the Phaethon surface were visible in two approaches during observations. Actually, in 2016 the asteroid showed the edge-on region of the surface (the mean aspect angle at the phase angle 106° – 85° was ~98°), while in 2017 the north-pole region (mostly covering the northern hemisphere and equatorial region of Phaethon) was seen (aspect angle at $\alpha \approx 62° - 135°$ was ~72°) (see Shinnaka et al. 2018; Okazaki et al. 2020). The difference in scattering properties of regolith of different regions of Phaethon's surface may provide large variations of polarization degree over the surface at large phase angles. McLennan et al. (2022) also detected the variations in thermal inertia across the Phaethon surface. For explanation, the authors proposed a two-component hemispherical model, in which Phaeton's northern and southern hemispheres have different thermophysical properties due to different material covering a northern (coarse-grained regolith) and southern hemispheres (fine-grained regolith).

To determine the parameters of the phase-angle dependence of polarization, such as the minimum polarization $P_{min}$ and minimum phase angle $\alpha_{min}$, the inversion angle $\alpha_{inv}$, the polarimetric slope $h$, the maximum polarization $P_{max}$ and the phase angle $\alpha_{max}$ at which this polarization is reached, several empirical approximations were used. Close to the opposition ($0° \leq \alpha \leq 30°$), a three-parameter (Kaasalainen et al. 2003) and four-parameter (Muinonen et al. 2009) relations derived from semiempirical modeling can be used for calculation of the parameters $P_{min}$, $\alpha_{min}$, $\alpha_{inv}$, and $h$. In a wider range of phase angles, the trigonometric function suggested by Lumme & Muinonen (1993) is usually applied. Shestopalov (2004) suggested another expression which contains five free parameters. However, all these empirical functions provide a best fit only in the case of good phase-angle coverage. In the case of Phaethon, the approximation using above mentioned expressions gives only very approximate values of the parameters of the phase-angle dependence of polarization, especially $P_{max}$ and $\alpha_{max}$. Since there are no polarization measurements of Phaethon at the negative polarization branch, we tried to use the data for asteroid (2) Pallas, because Phaethon may be its fragment (de León et al. 2010; Alí-Lagoa et al., 2013, 2016). However, as Fig. 2 demonstrates, these two asteroids have a distinct difference in polarization dependences. For example, $\alpha_{inv}$ for Phaethon is larger than that of Pallas. Note also that Kareta et al. (2018) and McLennan et al. (2022) both argue that Phaethon is an insufficient match to Pallas on dynamical and spectroscopic grounds.

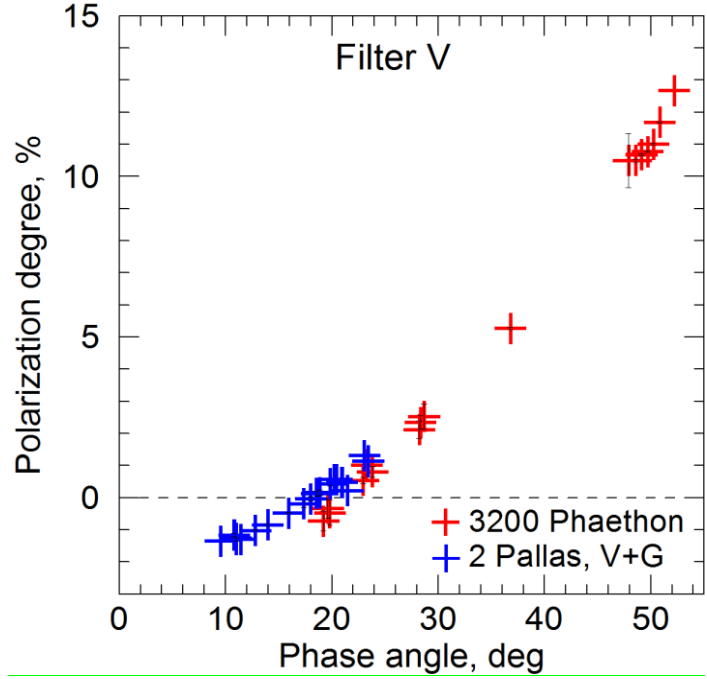

**Figure 2.** The polarization degree vs phase angle for (3200) Phaethon and (2) Pallas (Lupishko 2019) in the V filters. The data for Phaethon correspond to the data in Fig. 1.

In order to expand the range of phase angles, we combined the measurements obtained in the V and R filters. As seen in Fig. 3, the data for both filters are in good agreement within the measurement errors. This conclusion is supported by the data in the V and R bands reported by Devogéle et al. (2018). Usually, the expressions for data approximation assume that the positive polarization branch is symmetric. However, real objects often have a rather asymmetric branch of positive polarization (Levasseur-Regourd et al. 2015). To describe this asymmetry and incomplete data coverage, we modified the expression suggested by Shestopalov (2004):

$$P(\alpha) = \frac{h(1-e^{-m\alpha})(1-e^{-n(\alpha-\alpha_{inv})})(1-e^{-l(\alpha-\pi)})}{n(1-e^{-m\alpha_{inv}})(1-e^{-l(\alpha_{inv}-\pi)})} \times \frac{1}{(1+e^{q(\alpha-\alpha^*_{max})})}, \qquad (1)$$

which contains seven free parameters, $m$, $n$, $l$, $h$, $\alpha_{inv}$, $q$, and $\alpha^*_{max}$. Two new free parameters $q$ and $\alpha_{max}$ are added to the empirical formula by Shestopalov. The $q$ parameter determines the degree of asymmetry of the positive polarization branch, while the parameter $\alpha^*_{max}$ affects the position of polarization maximum. The solid curve in Fig. 3 is the fit to the polarization data of the asteroid Phaethon in the V and R filters using Eq. (1) with the following parameters: $m = -0.0142 \pm 0.0002$; $n = 0.0240 \pm 0.0003$; $l = -0.00364 \pm 0.00022$; $h = 0.196 \pm 0.002$; $\alpha_{inv} = 20.16° \pm 0.21°$; $q = 0.355 \pm 0.018$; $\alpha^*_{max} = 137.28° \pm 0.15°$.

Even though the apparent visual magnitude of Phaethon was weaker than $17^m$ and the asteroid was observed at the air mass ~4.2, we were able to obtain a rough estimate of the degree of polarization $P = 32\% \pm 3\%$ at a phase angle of $\alpha = 134.9°$ that allowed us to clarify the value and position of the maximum degree polarization of Phaethon. As a result, we have found that the maximum degree of polarization for Phaethon is $P_{max} = (45 \pm 1)\%$ at phase angle $\alpha_{max} = (124.0 \pm 0.4)°$. These parameters differ significantly from the values of $P_{max} = 48.6\%$ and $\alpha_{max} = 136.6°$ obtained in the V filter by Golubeva, Shestopalov & Kvaratskhelia (2020) using the data by Devogéle et al. (2018) and the approximation by Shestopalov (2004).

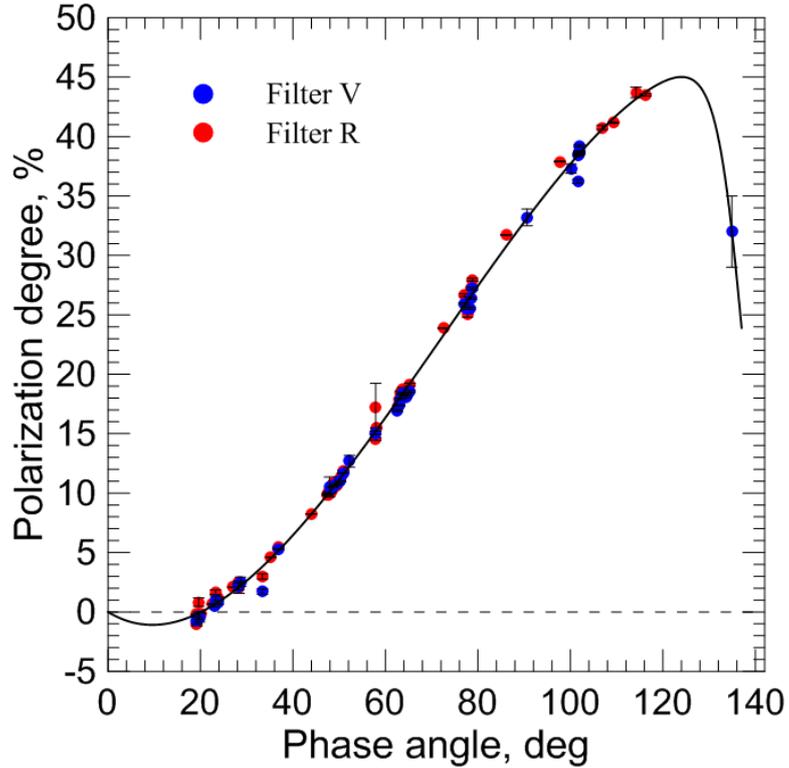

**Figure 3.** The polarization degree vs phase angle for (3200) Phaethon in the V and R filters. The data for Phaethon correspond to the data in Fig. 1. The black solid curve is a result of the least-square fit by applying Eq. (1).

*3.2. Determination of albedo of Phaethon*

The polarimetric method of determining the geometric albedo of asteroids $p_v$ is based on the empiric correlation between the albedo and the parameters describing the phase-angle dependence of polarization: albedo vs the minimum polarization degree $P_{min}$ of the negative polarization branch; albedo vs the polarimetric slope $h$ of the polarization curve determined around the inversion phase angle $\alpha_{inv}$; albedo vs the maximum value of the positive polarization branch $P_{max}$. These simplest methods require no assumptions, models, or simultaneously obtained photometric data. However, it requires the generalized calibration of the polarimetric albedo scale of asteroids. Since we do not have enough data on the negative branch of the phase-angle polarization curve for Phaethon and the exact position of the value of maximum polarization, we consider the relationship between geometric albedo and polarimetric slope $h$ (Zellner & Gradie 1976).

To estimate the inversion angle $\alpha_{inv}$, at which the polarization of asteroid Phaethon changes its sign, and the polarimetric slope $h$ of the ascending polarization branch, we analyzed our polarimetric data in the V band at phase angles 19.21° – 28.67° as well as all available data near inversion point, namely, Fornasier et al. (2006) at $\alpha = 22.94°$, Devogèle et al. (2018) at $\alpha = 36.79°$, and Okazaki et al. (2020) at $\alpha = 33.4°$. As can be seen in Fig. 4, the degree of polarization measured by Okazaki et al. is very different from other measurements, so their data were not taken into account in the further analysis. The measurements by Devogèle et al. were taken quite far from the inversion angle. Moreover, the polarization degree obtained at this phase angle may not correspond to the linear dependence, since there is the curvature of the ascending polarization branch in the region of phase angles greater than 35°. Therefore, we analyzed only our and Fornasier et al. polarimetric data, without the measurements by Okazaki et al. and Devogèle et al. (Fig. 4). A linear fit was applied and the best-fit parameters were derived: the polarimetric slope $h = (0.326 \pm 0.027)$

%/deg and the inversion angle $α_{inv}$ = 21.4° ± 1.2°. The computed $α_{inv}$ is consistent with that obtained by Shinnaka et al. (2018) in the R filter, but there is a rather large discrepancy with the value $α_{inv}$ = 18.8°±1.6° obtained by Devogéle et al. (2018) that can be explained by the lack of their data near the inversion point.

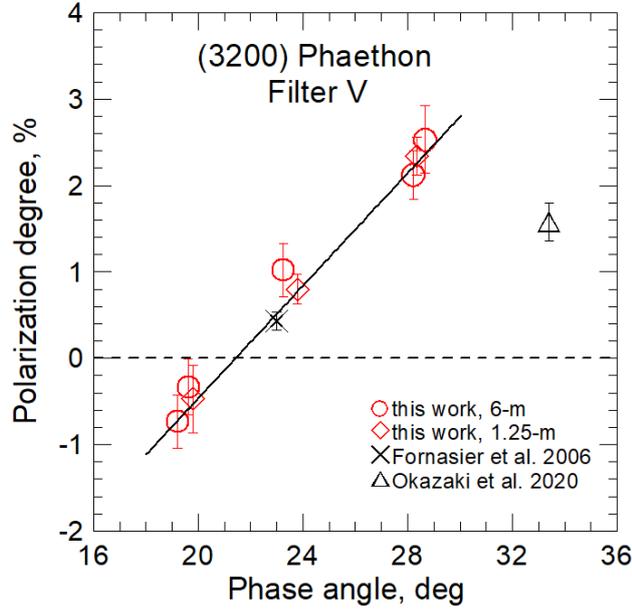

**Figure 4.** The best linear fit to the available data in the V filter within the range of phase angles around the inversion phase angle $α_{inv}$, from 19° to 29°. The black solid line is the fit to our data (red symbols) and Fornasier et al. (2006) (black oblique crosses). The measurement by Okazaki et al. (2020) (black triangle) is not included to the data fit.

Since the polarimetric slope $h$ directly (phenomenologically) depends on the maximum polarization $P_{max}$, there is also a close correlation between the geometric albedo $p$ and the polarimetric slope $h$. It is represented by the regression equation (KenKnight, Rosenberg, & Wehner 1967; Widorn 1967):

$$\log p = -C_1 \log h + C_2, \qquad (2)$$

where $C_1$ and $C_2$ are constants. For calculating albedo, we used empirical calibration coefficients determined by Lupishko (2018), namely $C_1$ = –(1.016 ± 0.010) and $C_2$ = –(1.719 ± 0.012). Using Eq. (2) and calculated polarimetric slope $h$ = (0.326 ± 0.027) %/deg, we estimated the geometric albedo of Phaethon, $p_v$ = 0.060 ± 0.005, which is consistent with that for dark asteroids: for example, the mean geometric albedo for C-type asteroids is $p_v$ = 0.061 ± 0.017 (Shevchenko et al. 2016). The obtained value of the geometric albedo is much smaller than the albedo $p_v$ = 0.122 derived by Hanuš et al. (2018) but is quite similar to other low values: 0.081 (Ito et al. 2018), 0.068 (Kareta et al. 2018), 0.06 (Taylor et al. 2019) (see Table 3).

Shinnaka et al. (2018) determined the polarimetric slope equal to $h$ = (0.174 ± 0.053) %/deg using observations in the Cousins $R$ filter within a wide range of phase angles, 19.2° – 87°. These authors did not provide the "albedo –slope" relation they used, but only the value $p_v$ = 0.14 ± 0.04. They noted that their albedo is close to the radiometric albedo 0.122 ± 0.008 obtained by Hanuš et al. (2018). However, as can be seen from Fig. 1, the polarization curve changes its slope, starting from phase angles of approximately 35° – 40° up to ~80°. Therefore, the phase-angle dependence of polarization cannot be represented by a single linear function in the range of phase angles 19° – 87°.

However, a linear fit to the Shinnaka et al. data points within the range of phase angles 19.2° – 35.2° provides the slope $h_R = 0.300 \pm 0.045$ %/deg and the derived albedo $p_R = 0.065 \pm 0.002$ is consistent with our result (see Table 3).

Available polarimetric measurements (Figs. 1 and 3) show that Phaethon has a very high polarization at large phase angles. As it is known, high maximum polarization indicates that the geometric albedo of the object should be low: there is a physically grounded (Gehrels 1977) connection between geometrical albedo $p$ and the maximum degree of polarization $P_{max}$. It manifests itself in the phenomenological Umow's law (Umov 1905): the smaller albedo of the reflecting surface, the higher its degree of polarization. This effect is caused by the fact that multiple scattering of light, which causes depolarization of the light, is more effective on surfaces with high albedo than on surfaces with low albedo. It should be noted that it was found experimentally (Geake & Dollfus 1986) that this connection differs from a strict inverse correlation.

In Table 3, we summarize all available determinations of geometric albedo for Phaethon derived by different techniques. Previous, up to the 2017 apparition, determinations of Phaethon's albedo, based on IR observations, unveiled the visual geometric albedo of Phaethon varying from $p_v = 0.17 – 0.16$ (Veeder, Kowal & Matson 1984; Usui et al. 2011) to about 0.1 (Green et al. 1985; Tedesco et al. 1992, 2002, 2004; Hanuš et al. 2016) and the effective diameter to be within the range $D \approx 4.2 – 5.1$ km. Our measured value of the geometric albedo of Phaethon, as well as other polarimetric estimates of albedo of this asteroid, and also the radar albedo, is significantly lower than albedo derived from the IR data. The difference between polarimetric albedo and infrared can be caused by the model dependence of the latter, since it is based on the thermophysical models of asteroids. The surface of Phaethon can be greatly processed due to the close passage near the Sun and therefore the thermophysical properties of the Phaethon's surface can differ significantly from the standard model. In addition, Phaethon sometimes exhibits an active dust coma (Jewitt & Li 2010) and, therefore, its surface may change over time.

**Table 3**
Geometric albedo of asteroid (3200) Phaethon determined by different methods.

| $p_v$ | Method and reference |
|---|---|
| 0.17 | IR data, non-rotating thermal model, diameter asteroid $D = 4.7$ km (Veeder, Kowal & Matson 1984) |
| 0.08 | IR data, fast-rotating thermal model, $D = 6.9$ km (Veeder, Kowal & Matson 1984) |
| $0.11 \pm 0.02$ ($p_R$) | IR photometry (radiometric diameter $D = 4.7 \pm 0.5$ km), (Green et al. 1985) |
| 0.098 | IRAS albedo (Tedesco et al. 1992) |
| $0.11 \pm 0.01$ | IRAS database, STM model (Tedesco et al. 1992) |
| $0.1066 \pm 0.011$ | IRAS observations, $D = 5.10 \pm 0.2$ km (Tedesco et al. 2002, 2004) |
| $0.16 \pm 0.01$ | AKARI data, radiometric diameter $D = 4.17 \pm 0.13$ km (Usui et al. 2011) |
| $0.122 \pm 0.008$ | IRAS and Spitzer data, thermophysical model, $D = 5.1 \pm 0.2$ km (Hanuš et al. 2016, 2018) |
| $0.08 \pm 0.01$ | NIR data, thermophysical model (Kareta et al. 2018) |
| 0.081 | Combination of radar diameter and photometry data (Ito et al. 2018) |
| $0.14 \pm 0.04$ | Polarimetry, slope-albedo relation (Shinnaka et al. 2018) |
| $(0.065 \pm 0.002)$ | (the value revised in this work) |
| $0.075 \pm 0.007$ ($p_v$) | Polarimetry (Zheltobryukhov et al. 2018) |
| ~0.05 | Polarimetry (Devogèle et al. 2018) |
| ~0.06 | Radar albedo (Taylor et al. 2019) |
| $0.122 \pm 0.03$ | Photometry, $D = 4.90 − 0.54/+0.79$ km (Lin et al. 2020) |
| $0.1066 \pm 0.011$ | JPL Small Body Database (JPL777), $D = 6.25 \pm 0.15$ km, radar data |
| $0.060 \pm 0.005$ | Polarimetry, slope-albedo relation, $D = 6.88 \pm 0.29$ km (this work) |

*3.3. Spectral dependence of polarization*

For studying the spectral dependence of polarization, we used data in the B, V, R, and I filters by Devogèle et al. (2018) obtained within the range of phase angles 37° – 79°. The phase dependence of the polarization in each filter was taken into account. For this, we calculated the degree of polarization in each band at phase angles 40° – 80° with the step of 5°, using Eq. (1). Fig. 5 shows the normalized spectral dependence of the polarization degree relative to the polarization in the V band, $P$(filter)/$P$(V). It can be seen that the degree of polarization increases noticeably between the B and I bands. As a consequence, the mean spectral slope of polarization $\Delta P/\Delta\lambda = 4.77 \pm 2.05$ %/nm for Phaethon between the B and I bands is positive at phase angles 40° – 80°, which is consistent with the data for C-type asteroids. According to Belskaya et al. (2009), the spectral slope of polarization for the low albedo C-type asteroids of the Main Belt is positive, while asteroids of S and M types are characterized by the negative values of the spectral slope, i.e. $\Delta P/\Delta\lambda < 0$. For comparison, we note that most comets, except for a few cases, show the increase of the polarization degree with increasing the wavelength in the visible and red domains of spectra (Kiselev et al. 2015).

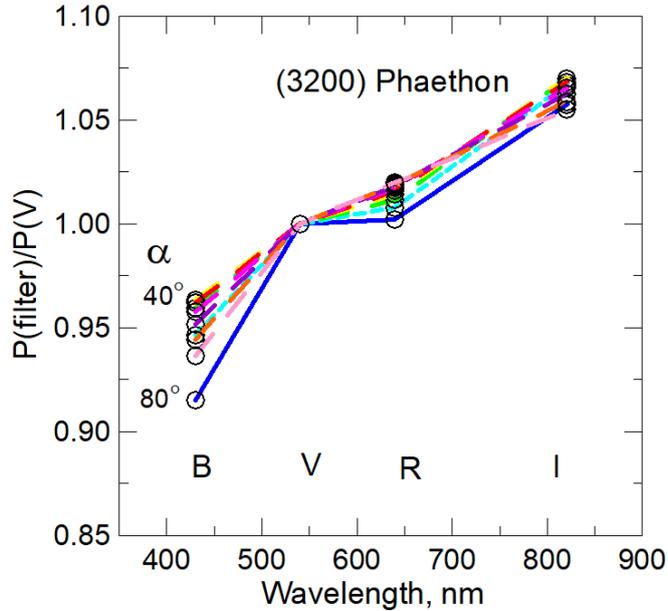

**Figure 5.** The slope of the normalized spectral dependence of polarization for Phaethon within the range of phase angles 40° – 80° (increment of 5°). Data were taken from Devogèle et al. (2018) for each band and corrected for the phase dependence of polarization according to Eq. (1).

## 4. Modeling of the phase-angle dependence of polarization

The small perihelion distance of Phaethon (0.14 au) could affect its surface composition and structure due to thermal heating by solar radiation. Licandro et al. (2007) concluded that Phaethon is more likely an active asteroid, similar to the population of active asteroids in the Main Belt, than an extinct comet. According to Delbo et al. (2014), thermal fragmentation induced by the diurnal temperature variations breaks up rocks larger than a few centimeters and releases dust that is carried away by the radiation pressure of sunlight. Outgoing dust particles are about a millimeter across, approximately the size of the Geminids. It is likely that as a result of thermal cracking in the Phaethon regolith, smaller (micron-size) particles can be also produced. Thermal fragmentation can affect not only the particle size of Phaethon's regolith, but also the composition and shape of the surface particles. Very strong heating of the asteroid surface obviously removes any kinds of water ice. Near-infrared observations of Phaethon reveal no evidence for hydration (Takir et al. 2020), while according to Licandro et al. (2007), the surface layer of Phaethon can be composed of hydrated silicates. Anyway, space weathering produces carbonization of the surface layers and, hence, carbon in the mixtures is reasonably expected.

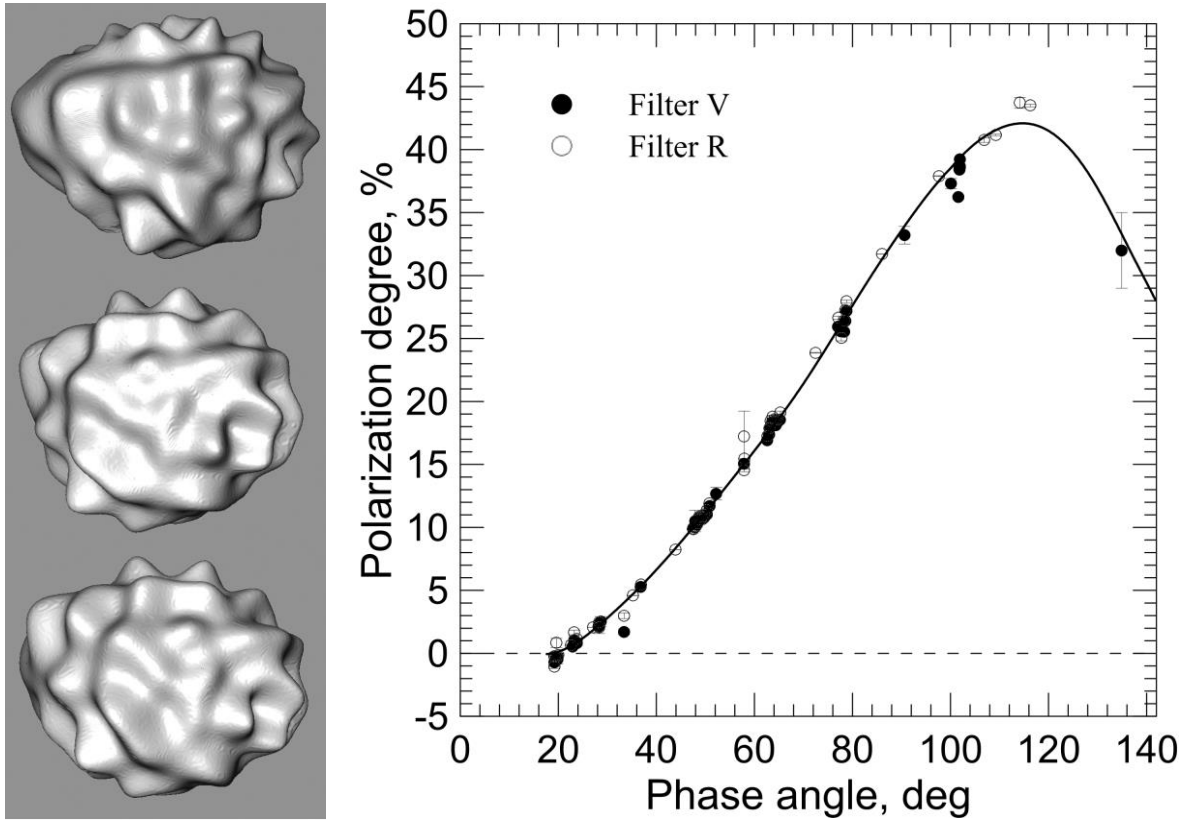

**Figure 6.** Left panel: The examples of Conjugated Random Gaussian particles (Petrov & Kiselev 2019). Right panel: Observational data for asteroid (3200) Phaethon in the V (solid circles) and R (open circles) filters. The black solid curve is the best fit of simulated polarization curve to the data by the mixture of 90% Mg-rich silicate and 10% carbon irregular particles.

To reveal the regolith properties of Phaethon via reproducing its phase-angle curve of polarization, the Sh-matrix method (Petrov et al. 2006; Petrov, Shkuratov & Videen 2011; Petrov, Shkuratov & Videen 2012) was used. For numerical simulations of the scattering properties, the complex refractive index as well as the size distribution of particles in dominant surface materials are required. A mixture of Mg-rich silicate (forsterite) particles (Scott & Duley 1996) and amorphous carbon particles (Rouleau & Martin 1991) was used as a composition model. Conjugated Random Gaussian particles were used as a model of irregular particles (Petrov & Kiselev 2019) (see Fig. 6). The main feature of these particles is presence of both large-scale and small-scale surface roughness. The parameters of the scattering particles, such as the size distribution and the carbon/silicates ratio were selected in the way produce the best approximation to the observational data. The solid line in Fig. 6 (right panel) corresponds to the intimate mixture of 90% Mg-rich silicates and 10% carbon, with the following size distribution parameters: effective size $R_{eff}$ and effective variance $v_{eff}$ defined by Hansen & Travis (1974): $R_{eff}^{(sil)} = 0.98 \ \mu m$;

$\upsilon_{eff}^{(sil)} = 0.21$; $R_{eff}^{(carb)} = 0.81 \ \mu m$; $\upsilon_{eff}^{(carb)} = 0.15$. Note that a mixture of carbon and silicates was used earlier by Kolokolova et al. (2015), where it was shown that particles with a size parameter greater than 25 (that corresponds to a size of 2.5 μm in the R filter) do not produce the negative polarization. Zubko et al. (2015) showed that the contribution of large particles to light scattering is rather small. The size of our particles, which provides the best agreement with the observational data, is in good accordance with the results of these papers. Also note that the sizes of model particles determined by us aren't very different from the estimates by Jewitt & Lee (2010), who determined that the size of dust particles coming off Phaethon at the Main Belt distances is ~1 μm.

## 5. UBV photometry of Phaethon

As we mentioned in Section 2.2, we used good photometric conditions on December 12, 2017 in order to determine the magnitude and colors of asteroid Phaethon from polarimetric observations at the 1.25-m telescope. The mean magnitude in the V band is $V(r,\Delta,\alpha) = 10.708^m \pm 0.039^m$ and the mean color indices $U–B = 0.207^m \pm 0.053^m$ and $B–V = 0.639^m \pm 0.054^m$ of the asteroid were derived at distances $r = 1.077$ au and $\Delta = 0.102$ au and phase angle $\alpha = 23.78°$. The running average variations of the V magnitude and U–B and B–V colors are plotted as a function of time in Fig. 7. The random uncertainties in the brightness ($\sigma = 0.02^m$) and color U–B ($\sigma = 0.03^m$) and B–V ($\sigma = 0.02^m$) are shown by the vertical lines. The amplitude of the lightcurve of Phaethon in the V band is about of $0.07^m$, while the variations of the U–B color index reach 0.7, and B–V about $0.1^m$. Thus, these values are within $3\sigma$ random uncertainties. This is probably a statistical noise, not an evidence of the own behavior of Phaethon. According to Kim et al. (2018), the lightcurve amplitude of Phaethon is $0.075^m \pm 0.035^m$, which is regarded as a nearly spherical shape.

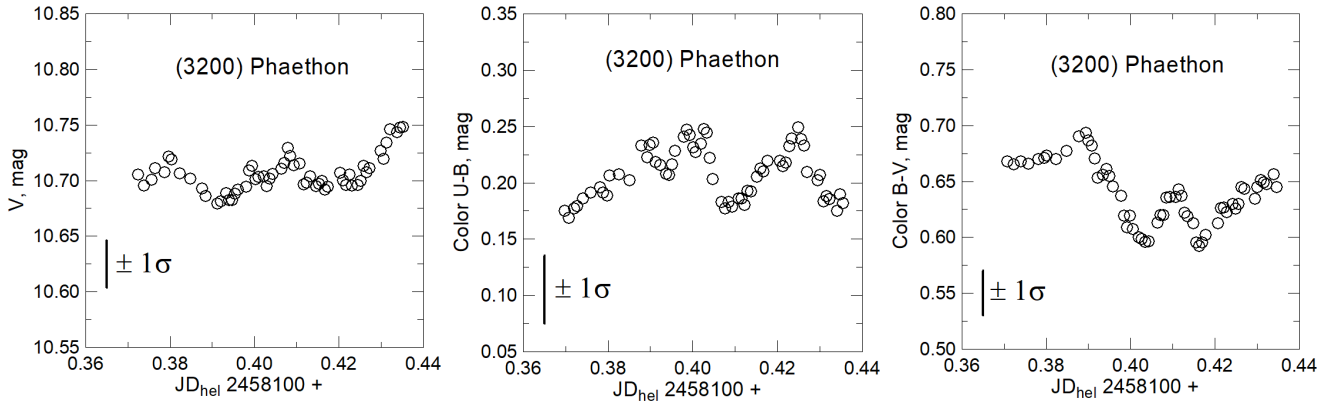

**Figure 7.** The lightcurve in the V band and the U–B and B–V colors variations of (3200) Phaethon on December 12, 2017. The running average data are presented. The random uncertainties in the brightness and colors are shown by the vertical lines.

The absolute magnitude reduced to unit heliocentric and geocentric distances and zero phase angle is defined as $V(1,1,0) = V(r,\Delta,\alpha) - 5\log 10(r\Delta) - \beta\alpha$, where $\beta$ is the linear phase coefficient of brightness. Using the linear coefficient brightness $\beta = 0.0422 \pm 0.0005$ mag/deg in the V band determined by Lin et al. (2020), we calculated the absolute magnitude of Phaethon to be $V(1, 1, 0) = 14.505^m \pm 0.059^m$. This value is based on linear regression which ignores the brightness opposition effect (BOE), which can vary within the range approximately $0.01^m – 0.4^m$, depending on the taxonomic class of asteroid (Shevchenko 2016). According to previous observations, the absolution magnitude of Phaethon is determined with relatively large uncertainty. For example, Luu & Jewitt (1990) derived for Phaethon $H = 14.65^m$. Latter Wisniewski et al. (1997) determined $H = 14.51^m \pm 0.14^m$, however after that Hanuš et al. (2016) corrected their $H$ value and provided the absolute magnitude $H = 14.41^m \pm 0.06^m$. According to the IRAS data (Tedesco et al., 2002), the absolute magnitude of Phaethon is $H = 14.51^m$. In the 2017 apparition, Lin et al. (2020) used the linear phase function to obtain the asteroid's absolute magnitude $H = 14.450^m \pm 0.017^m$.

Available in the literature colors *B–V*, *V–R*, and *R–I* of asteroid Phaethon, together with our *U–B* and *B–V* colors, are summarized in Table 4. These values are corrected for the influence of the solar color which are also given in the table. As seen, the color indexes of the asteroid show a scatter at the level of ~0.1$^m$. The mean colors of Phaethon derived from our observations are also given in the table together with the relevant uncertainties which are the standard errors of the mean. Our result for $B–V = 0.639^m \pm 0.054^m$ agrees with published colors for Phaethon within the uncertainties. The color index $U–B = 0.206^m \pm 0.043^m$ is firstly obtained for Phaethon. Most published optical colors of Phaethon are only slightly bluer than the Sun. We computed a weighted mean from the estimates given in Table 4. In general, with the exception of *U–B* which is slightly redder, the mean colors *B–V*, *V–R*, and *R–I* are slightly bluer than solar (Holmberg et al. 2006) that is consistent with the original classification of Phaethon as a F-type or B-type asteroid (Tholen 1985a,b; Green et al. 1985). The range of color values may indicate the composition heterogeneity of the Phaethon's surface. Comparing color indices with the existing data for Phaethon and also taking into account analysis by Dandy, Fitzsimmons & Collander-Brown (2003), Lee et al. (2019), and Lin et al. (2020), we concluded that asteroid Phaethon belongs to the B-type asteroids.

**Table 4**
Color indices for asteroid (3200) Phaethon.

| Date (UT) | *U–B* (mag) | *B–V* (mag) | *V–R* (mag) | *R–I* (mag) | Reference |
|---|---|---|---|---|---|
| 2017 Dec 12 | 0.21±0.05 | 0.64±0.05 | – | – | This work |
| Unknown | | | 0.34 | – | Skiff et al. (1996) |
| 1995 Jan 4 | | 0.52±0.01 | 0.33±0.01 | – | Ansdell et al. (2014) |
| 1996 Nov 12 | | 0.618±0.005 | 0.347±0.004 | – | Dundon (2005) |
| 1997 Nov 12 | | 0.58±0.01 | 0.34±0.01 | – | Ansdell et al. (2014) |
| 1997 Nov 22 | | 0.57±0.01 | 0.36±0.01 | – | Ansdell et al. (2014) |
| 1997 Nov 22 | | 0.650±0.004 | 0.295±0.002 | 0.320 ± 0.003 | Dundon (2005) |
| 2004 Nov 19 | | 0.587±0.005 | 0.349±0.003 | | Dundon (2005) |
| 2007 Sep 4 | | 0.61±0.01 | 0.34±0.03 | 0.27±0.04 | Kasuga & Jewitt (2008) |
| 2010 Sep 10 | | 0.67±0.02 | 0.32±0.02 | | Jewitt (2013) |
| 2003 Nov 14, 2004 Dec 20 | | – | 0.331 | – | Hanuš et al. (2016) |
| 2017 Oct 28 | | 0.64±0.02 | 0.34±0.02 | 0.31 ± 0.03 | Lee et al. (2019) |
| 2017 Oct 28 | | 0.69±0.07 | 0.35±0.04 | – | Kartashova et al. (2019) |
| 2017 Oct 29 | | 0.66±0.13 | 0.36±0.06 | – | Kartashova et al. (2019) |
| 2017 Nov 15 | | – | 0.43±0.02 | – | Kartashova et al. (2019) |
| 2017 Nov 17 | | 0.59±0.11 | 0.43±0.07 | | Kartashova et al. (2019) |
| 2017 Nov 22 | | 0.62±0.02 | 0.35±0.02 | – | Kartashova et al. (2019) |
| 2017 Nov 23 | | 0.64±0.04 | 0.36±0.03 | – | Kartashova et al. (2019) |
| 2017 Nov 27 | | 0.65±0.02 | 0.33±0.02 | – | Kartashova et al. (2019) |
| 2017 Dec 02 | | 0.65±0.01 | 0.35±0.01 | – | Kartashova et al. (2019) |
| 2017 Dec 13 | | 0.63±0.01 | 0.35±0.02 | – | Kartashova et al. (2019) |
| 2017 Dec 11–19 | | 0.702±0.004 | 0.309±0.003 | 0.266±0.004 | Tabeshian et al. (2019) |
| | | 0.633±0.036 | 0.336±0.011 | 0.334 ± 0.015 | Lin et al. (2020) |
| Average color | 0.21±0.05 | 0.627±0.044 | 0.331±0.006 | 0.296±0.015 | |
| Solar color | 0.173 ± 0.064 | 0.642 ± 0.016 | 0.354 ± 0.010 | 0.332 ± 0.008 | Holmberg et al. (2006) |

Using the relationship between the size, geometric albedo, and absolute magnitude of an asteroid, we computed the effective diameter $D$ of Phaethon following the common expression from (Bowell et al. 1989):

$$\log p_V = 5.642 - 0.4 m_V(1,1,0) - 2 \log R \qquad (3)$$

where $R$ is the asteroid's radius in kilometers. As an absolute magnitude $m_V(1,1,0)$, different authors use either the value $V(1,1,0)$ or the value $H(1,1,0) = V(1,1,0) - $ BOE, where BOE is brightness opposition effect. According to the definition, geometric albedo of a celestial body is the ratio of its actual brightness at zero phase angle to that from a flat Lambertian disk of the same cross-section. The brightness depends on the projection area of the illuminated part of the asteroid and on the change in the average brightness of the illuminated part of the disk, caused by the effects of light scattering (diffuse and coherent backscattering) in the surface layer of the asteroid. Since only diffuse scattering is considered for the Lambertian disk, the geometric albedo can numerically exceed 1. From a physical point of view, the value $V(1,1,0)$ should be taken as the absolute magnitude for the determination of the asteroid's diameter in Eq. (3). Using $V(1,1,0) = 14.505^m$ and geometric albedo $p_v = 0.060$, we derived the effective diameter of Phaethon $D = 6.8 \pm 0.3$ km. The error in the diameter was calculated using the uncertainty in the absolute magnitude and the albedo. According to Taylor et al. (2019), the radar images of the asteroid show a roughly spheroidal shape with a diameter of 6.2 km, whereas the diameter of Phaethon published in the ALCDEF database (http://alcdef.org/) is $5.8 \pm 0.3$ km Usually, the radiometric sizes are systematically lower, e.g., $5.1 \pm 0.2$ km (Tedesco et al. 2002) and $4.17 \pm 0.13$ km (Usui et al. 2011). The diameter of Phaethon defined from observations of star occultation by the asteroid is 5.2 km (Ye et al. 2019). The same estimate of the effective diameter ($5.2 \pm 0.1$ km) was obtained by Devogèle et al. (2020).

The brightness amplitude of NEAs depends on the axis ratio and the aspect angle $\psi$ of the asteroid, that is the angle between the rotation axis of the asteroid and the line of sight, during the observations. We determined the average aspect for Phaethon equal to $\psi \approx 68°$ at the time of our photometric observations, using the formula by Zappala (1981):

$$\cos \psi = -(\sin \beta \sin \beta_0 + \cos \beta s \cos \beta_0 \cos(\lambda - \lambda_0)), \qquad (4)$$

where $\beta$ and $\lambda$ are ecliptic latitude and longitude of the asteroid; ecliptic coordinates of the pole $\beta_0 = -47° \pm 5°$ and $\lambda_0 = 318° \pm 5°$ are taken from Hanuš et al. (2018). Since the lightcurve amplitude varies as a function of the polar aspect viewing angle $\psi$ (Binzel et al. 1989), we can estimate lower limit of the axis ratio $a/b$, having the amplitude of the lightcurve, from the expression $\Delta V(\psi = 90°) = 2.5 \log(a/b) \approx \Delta V_{max} \approx 0.08^m$, where $a$ and $b$ are, respectively, the long and short axis of an elongated body. As a result, the ratio axis $a/b$ for Phaethon projected on the sky plane is 1.08 that indicates that the asteroid is nearly spheroidal in shape. This result is consistent with the image of Phaethon from radar image derived by Taylor et al. (2019). Hanuš et al. (2018) showed that the overall shape of Phaethon is nearly axially symmetric with $a/b \approx 1.06$. Kartashova et al. (2019) found that Phaethon's maximum amplitude in the V filter was $0.11^m$ which corresponds to the axes ratio $a/b = 1.11$, and Lin et al. (2020) obtained the same values for the amplitude and the ratio a/b.

## 6. Reflection spectrum

Together with polarimetry and multicolor photometry, we carried out a visible spectrometry to investigate the changes in the spectra of Phaethon over time. The spectra of asteroid Phaethon were obtained from spectropolarimetric measurements with the SCORPIO-2 instrument during six nights, from November 18 to December 13, 2017 (see Table 1). We used the specialized software packages in the IDL environment developed in the SAO RAS and standard procedures to reduce the obtained data (see for details Afanasiev & Amirkhanyan (2012) and Ivanova et al. (2017)). The reflectance spectra of Phaethon normalized to the average intensity in the range $\lambda$5500–6000Å are presented in Fig. 8. In this figure, we also display standard solar spectrum (Neckel & Labs 1984), which was transformed to the SCORPIO-2 resolution by convolving with instrumental profile and also normalized to the average intensity within the range $\lambda$5500–6000Å. We checked the asteroid spectra for their absorption lines. For this, we compared the solar spectrum with the solar absorption features seen in the spectra of the asteroid. As can be seen in these figures, the absorption features correspond to the major Fraunhofer lines, and the Phaeton's spectra are featureless at wavelengths longer than $\lambda$4600 Å. In all spectra, the reflectance decreases below $\lambda$4600 Å. Also, there are significant differences among the spectra within the range of wavelengths from approximately $\lambda$4600 Å to $\lambda$8000 Å. The reflectance spectra obtained on November 27 and 28 are very similar and nearly coincide with the solar spectrum. On December 9.001, the spectrum has changed dramatically and had a negative slope, starting at wavelengths of about $\lambda$5000. But on December 9.754, the spectrum was in general agreement with the spectra taken on November 27 and 28. Phaeton's spectrum changed even more on December 12 and 13. Such differences in the spectra may indicate the inhomogeneity of surface composition of Phaeton.

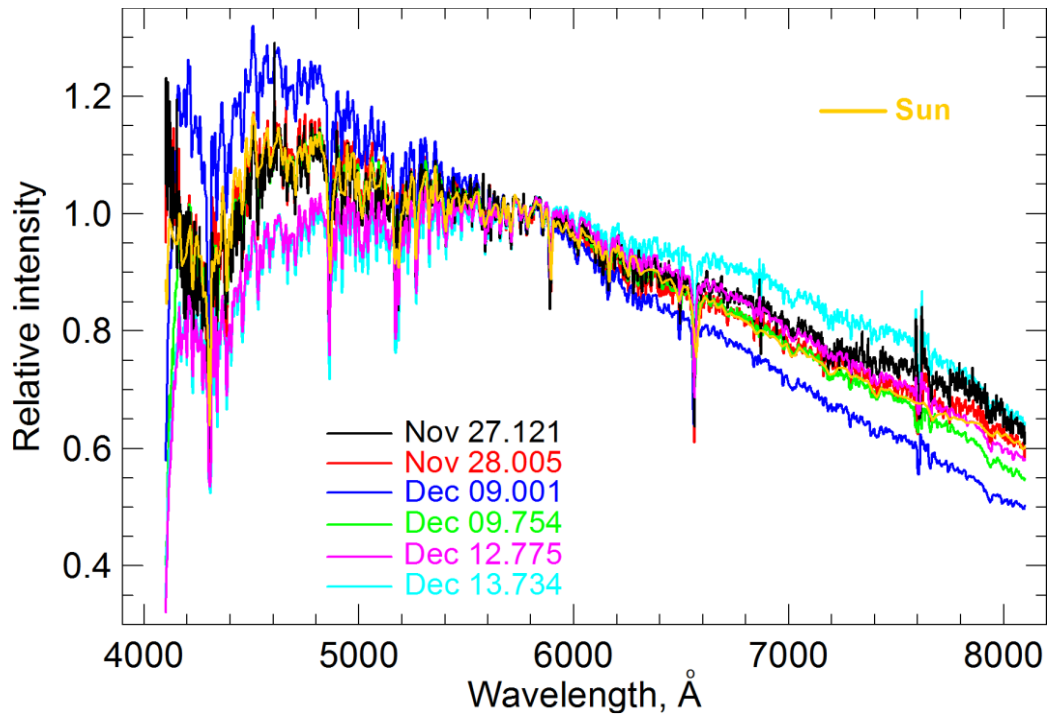

**Figure 8.** Reflectance spectra of asteroid Phaethon obtained within the period of observations from November 27 to December 13, 2017 are shown together with the solar spectrum. The color code represents the different observing dates. For comparison, the normalized spectrum of the Sun marked by yellow line is shown. All spectra are normalized to the average intensity in the range $\lambda$5500–6000Å. Spectra are shown chronologically from top to bottom. The absorption features correspond to the major Fraunhofer lines.

We note one more feature of the obtained Phaethon spectra, namely, a difference in the position of the reflectance maximum (see Fig. 8). From November 27 to December 9.75, the maximum is located at about λ4600 Å, while on December 12 and 13 a maximum is near λ5000 Å. Since the detected spectral variations are larger than the observational uncertainties, we consider them real. A similar spectral behavior has been noted by Licandro et al. (2007).

The smoothed ratio of the observed spectrum of the asteroid to the solar spectrum is shown in Fig. 9. To calculate the spectral slopes, we applied the linear least-squares fit to the Phaethon's spectra in the wavelength range λ4600–7500 Å. The normalized spectral gradient of reflectivity $S'$ between wavelengths $\lambda_1$ and $\lambda_2$ (in units of %/1000 Å) was computed as in Luu & Jewitt (1990): $S'(\lambda_1,\lambda_2) = (dS/d\lambda)/S_c$, where $dS/d\lambda$ is the rate of change of the reflectivity with respect to the wavelength within the range between $\lambda_1$ and $\lambda_2$, and $S_c$ is the average reflectivity in the range λ6000–6100Å. The apparent (uncorrected for phase) spectral gradient $S'$ and its error $\sigma S'$ are given in Table 5. This table also includes the date and midtime of the observation, the heliocentric $r$ and geocentric $\Delta$ distances, the phase angle $\alpha$, and the aspect angles (i.e., sub-Earth latitude on the asteroid) $\psi$. The total slope error consists of the error in the linear fit (0.001–0.002 %/1000 Å) and the uncertainty introduced by the calculated intensities after dividing Phaethon's spectra by the solar spectrum which is dominant in the final error.

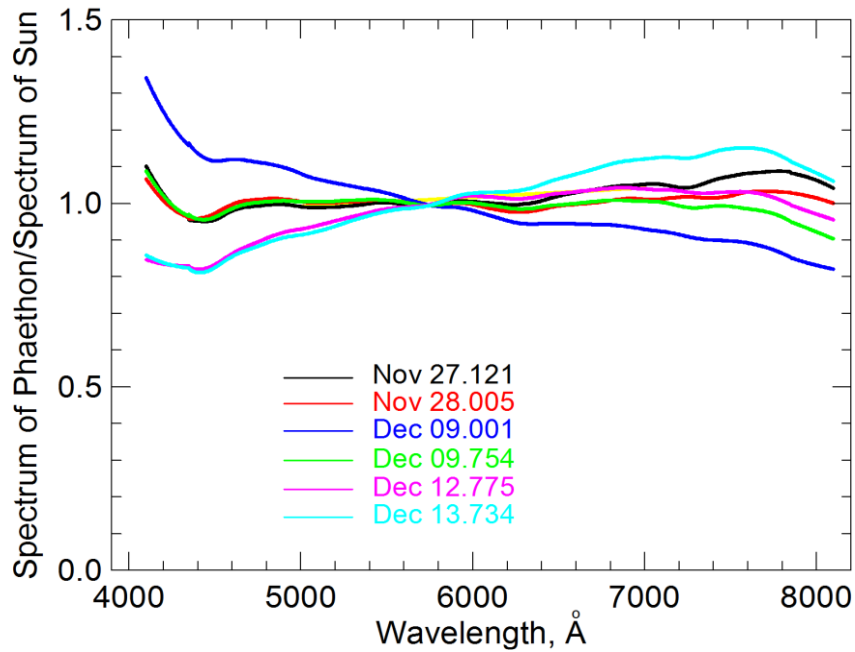

**Figure 9.** Smoothed ratio of the asteroid Phaethon spectrum to the solar spectrum. Spectra are normalized at average intensity within the range λ5500–6000 Å. The color code reflects the different observing nights.

The computed spectral slopes, obtained for the period between November 27 and December 13, 2017, are obviously time-variable, from –7.78 to 9.18 %/1000 Å (see Table.5). The Phaethon spectra were well studied by many observers (e.g., Tholen 1985b; Luu & Jewitt 1990; Luu 1993; Lazzarin et al. 1996; Katsuhito et al. 2020; Kareta et al. 2018) and in all cases the spectra exhibited large variations in the spectral gradient. In the 2017 apparition, the slightly blue reflectivity gradient $S' = -(2.43 \pm 0.08)$ %/1000 Å in the visible spectrum (λ4000–7400 Å) was detected by Kareta et al. (2018). Very blue color $S' = -(13.3 \pm 1.0)$ %/1000 Å in the spectral range λ4800–7200 Å was found by Luu (1993). Our measurements are consistent with the slope values of spectra obtained by Ohtsuka et al. (2020). According to these authors, the spectral gradient of Phaethon varied between

−5.0 %/1000 Å and 0.6 %/1000 Å during its 2007 return. They noted that rotational modulation of the spectral slope of this asteroid suggests its surface compositional inhomogeneity. Previously, Licandro et al. (2007) also suggested that a very inhomogeneous surface of Phaeton was heavily altered due to different composition of materials on its surface caused by the thermal alteration at the close proximity of the Sun due to a very small perihelion distance $q \approx 0.14$ au, or/and by non-uniform heating due to Phaethon's pole orientation.

**Table 5**
The normalized reflectivity gradients within the wavelength range λ4800–7200 Å

| Date (2017, UT) | $r$ (au) | $\Delta$ (au) | $\alpha$ (deg) | $\Psi$ (deg) | $S' \pm \sigma S'$ (%/1000 Å) |
|---|---|---|---|---|---|
| Nov 27.121 | 1.305 | 0.382 | 28.67 | 75.54 | 2.27 ± 0.04 |
| Nov 28.005 | 1.293 | 0.365 | 28.22 | 75.31 | –0.15 ± 0.04 |
| Dec 09.001 | 1.137 | 0.163 | 19.61 | 72.02 | –7.87 ± 0.04 |
| Dec 09.754 | 1.126 | 0.150 | 19.21 | 71.76 | 0.03 ± 0.02 |
| Dec 12.775 | 1.079 | 0.104 | 23.24 | 70.66 | 5.24 ± 0.07 |
| Dec 13.734[a] | 1.064 | 0.091 | 28.34 | 70.62 | 9.18 ± 0.03 |

[a] partially cloudy.

## 7. Conclusions

We have presented results of polarimetric, photometric, and spectral observations of the near-Earth asteroid (3200) Phaethon carried out at the 6-m BTA telescope of the Special Astrophysical Observatory and the 2.6-m and 1.25-m telescopes of the Crimean Astrophysical Observatory during 2017–2020 at the phase angle 19° to 135° and their analysis. Our findings are the following:

1) The surface of asteroid Phaethon exhibits extremely large linear polarization. The phase-angle dependence of polarization in the V filter is characterized by the following parameters: $P_{max} = 45\% \pm 1\%$ at the phase angle $\alpha_{max} = 124.0° \pm 0.4°$, the inversion angle $\alpha_{max} = 21.4° \pm 1.2°$, and polarimetric slope $h = 0.326 \pm 0.027$ %/deg. In the previous 2016 apparition, Ito et al. (2018) revealed even higher polarization: $P = 50.0 \pm 1.1\%$ at $\alpha = 106.5°$. These results may be evidence for surface heterogeneity of Phaethon, which can manifest itself in the variations of the composition, particle sizes, or porosity of the regolith over the surface.

2) According to our polarimetric data, the geometric albedo of Phaethon is $p_v = 0.060 \pm 0.005$ which is significantly lower than the estimates obtained using other techniques (e.g., IR data).

3) The mean spectral slope of polarization for Phaethon at phase angles 40° – 80° is positive between the B and I bands, $\Delta P/\Delta \lambda = 4.77 \pm 2.05$ %/nm, which is similar to that for the low albedo asteroids.

4) The best-fit to the observational phase-angle curve of polarization, using the Sh-matrix model of conjugated random Gaussian particles, is obtained for a mixture of particles consisting of Mg-rich silicates (90%) and amorphous carbon (10%).

5) The absolute magnitude and the mean color indices of Phaethon, obtained with a relatively large uncertainty, are: $V(1,1,0) = 14.505^m \pm 0.059^m$, $U–B = 0.207^m \pm 0.053^m$, and $B–V = 0.639^m \pm 0.054^m$.

6) The effective diameter of Phaethon estimated from the obtained absolute magnitude and geometrical albedo, is $6.8 \pm 0.3$ km which is larger than previously reported size $5.1 \pm 0.2$ km from the thermophysical modeling by Hanuš et al. (2016), or the best-fit spherical diameter of Phaethon about 6.2 km from the Arecibo radar images (Taylor et al. 2019), or new estimate for the diameter 5.4 ± 0.5 km determined by Devogèle et al. (2020). The ratio axis *a*/*b* for Phaethon projected on the sky plane has been found to be 1.08 that confirms the conclusions of other authors that the asteroid is nearly spheroidal in shape.

7) The reflectance spectra taken over six nights, from November 18 to December 13, 2017, exhibited no absorption features and for different nights demonstrated different spectral slopes in the wavelength range $\lambda$4800–7200 Å, which varied between –7.78 to 9.18 %/1000 Å.

ACKNOWLEDGMENTS

The authors note the great contribution of Prof. Viktor Afanasiev to this work, who passed away on December 21, 2020. He obtained and partially processed the observations of asteroid Phaethon taken with the 6-m telescope of the SAO.

The authors appreciate Ludmilla Kolokolova for her insightful discussion. VK, IL, and OI are supported, in part, by the project 22BF023-02 of the Taras Shevchenko National University of Kyiv. The work by OI was supported by the Slovak Grant Agency for Science VEGA (Grant No. 2/0059/22) and by the Slovak Research and Development Agency under the Contract No. APVV-19-0072.

Data availability

The data underlying this article will be shared on reasonable request to the corresponding author.